\documentclass{ws-ijmpcs}
\usepackage{url,overcite}

\renewcommand{\draftnote}[1]{
}

\begin{document}
\markboth{S.C. Lim \& C.H. Eab}
{Some Fractional and Multifractional Gaussian Processes: A Brief Introduction}

%
%

\title{Some Fractional and Multifractional Gaussian Processes: A Brief Introduction}

\author{S.C. Lim}
\address{%
Faculty of Engineering, Multimedia University,
 63100 Cyberjaya, Selangor Darul Ehsan, Malaysia%
\\
{sclim47@gmail.com}
}

\author{Chai Hok Eab}
\address{%
Department of Chemistry,
Faculty of Science, Chulalongkorn University,
Bangkok 10330, Thailand%
\\
{Chaihok.E@chula.ac.th}
}

\maketitle


\begin{abstract}
This paper gives a brief introduction to some important fractional and multifractional Gaussian processes commonly used in modelling natural phenomena and man-made systems. The processes include fractional Brownian motion (both standard and the Riemann-Liouville type), 
multifractional Brownian motion, fractional and multifractional Ornstein-Uhlenbeck processes,
fractional and mutifractional Reisz-Bessel motion. Possible applications of these processes are briefly mentioned. 
\end{abstract}

\ccode{PACS numbers: 02.50.-r, 02.50.Ey, 05.40.-a}

\section{Introduction}
\label{sec:intro}

During the past few decades, fractional calculus%
\cite{OldhamSpanier74,MillerRoss93,Ortigueira11,Herrmann13}
has found applications in diverse fields ranging from physical and biological sciences, engineering to internet traffic and economics. 
One of the main reasons for its popularity in modelling many phenomena
is that it provides a natural setting for describing processes which are fractal in nature and with memory.%
\cite{Nigmatullin92,Rutman95,Tatom95,oshrefiTorbatiHammond98,Podlubny02,TenreiroMachado03,Stanislavsky04}%
Many applications of fractional calculus are based on the fractional integro-differential equations.%
\cite{Samko93,Podlubny99,Kilbas06,Diethelm10}%
For example, various types of fractional diffusion equations and fractional
Langevin-type equations have been proposed to model anomalous diffusion, and both
deterministic and stochastic fractional equations are used to describe viscoelastic phenomena,
telecommunication, and other systems in science and engineering.%
\cite{MetzlerKlafter00,Hilfer00,West03,Zaslavsky05,Klages08,Mainardi10,Das11,KlafterLimMetzler11,Tarasov11,%
Diethelm12,Sheng11,Meerschaert12,Uchaikin12,Atanackovic14}

The usual way to obtain concrete realization of a particular fractional model is to
associate it with a fractional generalization of an ordinary stochastic process. 
This can be carried out nicely due to the smooth integration of fractional calculus and probability
theory. 
The most well-known among these fractional stochastic processes include
fractional Brownian motion 
\cite{Biagini08,Mishura08,Nourdin12}
and fractional Levy motion.%
\cite{SamorodnitskyTaqqu94,Marquardt06,Cohen13}
Another fractional
stochastic process of interest is fractional Ornstein-Uhlenbeck process.%
\cite{Cheridito03,LimMuniandy03,Magdziarz08}

Fractional Brownian motion (FBM) and fractional Ornstein-Uhlenbeck (FOU)
process are characterized by a single parameter. 
It is possible to extend FBM to 
bifractional Brownian motion%
{\cite{Russo06}} 
and mixed FBM,%
{\cite{Cheridito01}}
which are indexed by two parameters and two or more parameters respectively.
Similarly, FOU process can also be generalized to a process parametrized by
two fractional indices.\cite{LimLiTeo08,LimTeo09}
Other examples of stochastic processes with two indices are fractional Riesz-Bessel motion (FRBM)
\cite{Anh99,AnhLeonenkoMcvinish01}
and Gaussian process with generalized Cauchy covariance (generalized Cauchy process).%
\cite{GneitingSchlather04,LimLi06}
In general, processes parametrized by two indices can provide more flexibility in 
modelling physical phenomena. 
In the case of the generalized Cauchy process both have the advantage that
the two indices provide separate characterization of the fractal dimension or self-similar
property, a local property, and the long-range dependence, a global property. 
This is in contrast to models based on fractional Brownian motion which characterizes these two
properties with a single parameter.
On the other hand, the two indices of FRBM characterize the long-range dependence
and intermittency separately.
In contrast, FBM is not intermittent.

Further generalization of fractional process can be carried out by replacing the constant index by a continuous function of time. In this way, one obtains multifractional
Brownian motion 
\cite{PeltierVehel95,BenassiJaffardRoux97}
and multifractional Ornstein-Uhlenbeck process.%
\cite{LimTeo07}
Similarly, it is possible to have the multifractional extension of Riesz-Bessel motion
\cite{RuizMedina03,LimTeo08}
and generalized Cauchy process. 
These processes can be used to describe systems with variable fractal dimension and variable memory.
In this short paper we shall restrict our discussion on some fractional and
multifractional Gaussian processes, and mentioned briefly their possible applications.
The non-Gaussian fractional and multifractional Levy motion will not be considered here.


\section{Fractional Brownian Motion}
\label{sec:fracBM}
Among all the fractional stochastic processes applied to modeling natural and man-made systems, 
fractional Brownian motion (FBM) can be regarded as the most widely used. 
Here we would like to summarise briefly the main properties of FBM. 

The standard FBM as introduced by Mandelbrot and Van Ness 
\cite{MandelbrotVanNess68}
 is defined by the following moving average representation: 
\begin{align}
  D_H(t) & = \frac{1}{\Gamma(H+1/2)}
             \Biggl\{
             \int_{-\infty}^0 \bigl[(t-u)^{H-1/2} - (-u)^{H-1/2}\bigr]dB(u)
\nonumber \\
& \hspace{3cm} +
             \int_{0}^t (t-u)^{H-1/2}dB(u)
             \Biggr\},
\label{eq:fracBM_0010}
\end{align}
where $B(t)$ is the standard Brownian motion, 
$\Gamma$ is the gamma function and the H\"{o}lder exponent (or Hurst index) 
$H$ lies in the range 
$0 < H < 1$. 
Equation (\ref{eq:fracBM_0010}) can be written more compactly as 
\begin{align}
  B_H(t) & = \frac{1}{\Gamma(H+1/2)}
             \int_{-\infty}^\infty \bigl[(t-u)_{+}^{H-1/2} - (-u_{+})^{H-1/2}\bigr]dB(u),
\label{eq:fracBM_0020}
\end{align}
where $(x)_{+} = \max(x,0)$. 
Note that there exists an equivalent representation of FBM
known as the harmonizable or the spectral representation:\cite{SamorodnitskyTaqqu94}
\begin{align}
  B_H(t) & = \frac{1}{2\pi}\int_{-\infty}^\infty \frac{e^{it\xi}-1}{|\xi|^{H+1/2}}dB(\xi).
\label{eq:fracBM_0030}
\end{align}
$B_H$ is a Gaussian process with zero mean and its variance and covariance are respectively
\begin{align}
  \Bigl<\bigl(B_H(t)\bigr)^2\Bigr> & = \sigma_H^2 |t|^{2H} ,  \label{eq:fracBM_0040}\\
  \Bigl<\bigl(B_H(t)B_H(s)\bigr)^2\Bigr> & = \frac{\sigma_H^2}{2}\bigl[ |t|^{2H} +  |s|^{2H} -  |t-s|^{2H}\bigr], \label{eq:fracBM_0050}\\
\intertext{with}
   \sigma_H^2 & = \Bigl<(B_H(1)\bigr)^2\Bigr> = \frac{\Gamma(1-2H)\cos(\pi{H})}{\pi{H}}. \label{eq:fracBM_0060}
\end{align}
FBM defined above is continuous everywhere non-differentiable with an unique scaling exponent $H$, 
a characteristic of a monofractal process. 

The standard FBM $B_H$ has some desirable properties. 
It is a self-similar process of  order $H$: 
\begin{align}
  B_H(at) & = a^H B_H(t), &
          & \forall a > 0, \quad t \in \mathbb{R},
\label{eq:fracBM_0070}
\end{align}
where the equality is in the sense of finite joint distributions. 
Though $B_H$ is itself non-stationary, its increment process 
\begin{align}
  \Delta{B_H(t,\tau)} &\equiv B_H(t+\tau) - B_H(t), & 
     \tau & > 0,
\label{eq:fracBM_0080}
\end{align}
is stationary with covariance 
\begin{align}
  \Bigl<\Delta{B_H(t,\tau_1)}\Delta{B_H(t,\tau_2)}\Bigr> & = \frac{\sigma_H^2}{2}\bigl[ |\tau_1|^{2H} +  |\tau_2|^{2H} -  |\tau_1-\tau_2|^{2H}\bigr] .
\label{eq:fracBM_0090}
\end{align}
Self-similarity together with stationary increments imply 
\begin{align}
  B_H(t+\tau) - B(t) & = a^{-H}\bigl[B_H(t+a\tau)-B_H(t)\bigr], & \forall a > 0, \quad t \in \mathbb{R}.
\label{eq:fracBM_0100}
\end{align}

In contrast to the local properties which depend mainly on the correlations between points that are close to each other, 
the long and short-range dependence of a stochastic process is a global property that measure the total strength of the correlation over a large domain. 
Given a Gaussian stochastic process $Y(t)$ with correlation  
$R(t,s)= \Bigl<Y(s)Y(t)\Bigr>\biggl[\Bigl<\bigl(Y(s)\bigr)^2\Bigr>\Bigl<\bigl(Y(t)\bigr)^2\Bigr>\biggr]^{-1/2}$
we say that it has long-range dependence (LRD) or long memory if the integral  
$\int_{\mathbb{R}}R(t,t+u)du$ is divergent. 
On the other hand, if the integral is convergent, the process has short-range dependence (SRD) or short memory.
\cite{LimMuniandy03}
One can easily verify that FBM is LRD except for $H = 1/2$, which corresponds to Brownian motion, a Markov process.

Despite of the nice properties mentioned above, the standard FBM does not represent a causal time-invariant system as there does exist a well-defined impulse response function. 
There is another type of FBM, the one-sided FBM first introduced by 
Barnes and Allan\cite{BarnesAllan66} using the Riemann-Liouville (RL) fractional integral: 
\begin{align}
  X_H(t) & = \frac{1}{\Gamma(H+1/2)}\int_0^t (t-u)^{H-1/2}dB(u),
\label{eq:fracBM_0110}
\end{align}
represents a linear system driven by white noise $\eta(t)$, 
with the impulse response function
$t^{H-1/2}/\bigl(\Gamma(H+1/2)\bigr)$.
 The RL-FBM $X_H(t)$ is a zero-mean Gaussian process with a complicated covariance:
 \begin{align}
    \Bigl<X_H(t)X_H(s)\Bigr> & = \frac{t^{H-1/2}s^{H+1/2}}{\bigl(H+1/2\bigr)\bigl(\Gamma(H+1/2)\bigr)^2}\,
                                 {_2F_1}\bigl(1,1/2-H,3/2+H,s/t\bigr),
\label{eq:fracBM_0120}
\end{align}
where $s < t$ and ${_2F_1}$ is the Gauss hypergeometric function. 
However, the variance of $X_H$ has the same time dependence as $B_H$:
 \begin{align}
    \Bigl<\bigl(X_H(t)\bigr)^2\Bigr> & = \frac{t^{2H}}{2H\bigl(\Gamma(H+1/2)\bigr)^2} .
\label{eq:fracBM_0130}
 \end{align}

Except for the absence of stationary increments, 
$X_H$ has many properties in common with $B_H$, such as self-similarity, regularity of sample path, LRD, etc. 
Absence of stationary property for its increments implies that $X_H$ can not have a harmonizable representation, 
and it is also not possible to associate
to $X_H$ a generalized spectrum of power-law type as in the case of standard FBM. 
This is the main reason for the lesser use of FBM of RL-type in modeling systems with power­ law type spectrum. 
However, $X_H$ has gained more popularity recently in some applications as the process is physically more realistic since it starts at time zero. 

Applications of FBM are well-known and diverse. Here we just mention the more common ones such as 
anomalous transport phenomena in physical and biological sciences, telecommunication,\cite{Sheluhin07}
finance,\cite{Biagini08,Rostek09}


\section{Multifractional Brownian Motion}
\label{sec:multifracBM}
FBM can only be used in modelling phenomena which are monofractal with same irregularity globally 
and with constant memory as characterised by the constant H\"{o}lder exponent $H$. 
However, for real world systems global self-similarity seldom exists. Fixed scaling only holds for a certain finite range of intervals. 
In addition, empirical data indicates that the scaling exponent or order of self-similarity usually has more than one value.
Thus in many complex heterogeneous systems there exist phenomena which exhibit multifractal properties with variable space and time dependent memory.%
\cite{KKK03a,KKK03b,Sun09}
One simple way to generalize a mono-scaling FBM to a multi-scaling FBM (or multifractional Brownian motion, MBM) is to replace the constant H\"{o}lder exponent by
$H(t), t \in \mathbb{R}^{+}$,
a $(0,1)$-valued function with H\"{o}lder regularity 
$r, r > \sup{H(t)}$. 
In general $H(t)$ can be a deterministic or random function, 
and it needs not be a continuous function. 
This time-varying H\"{o}lder exponent $H(t)$ describes the local variations of the irregularity of the MBM. 
Such a generalization of FBM $B_H$ to MBM $B_{H(t)}$ was carried out independently by 
Peltier and L\'{e}vy-V\'{e}hel\cite{PeltierVehel95}
based on the moving-average representation and by 
Benassi {\it et al}\cite{BenassiJaffardRoux97} 
using the harmonizable representation. As expected, these two generalizations of MBM are almost certainly equivalent up to a multiplicative deterministic function of time.%
\cite{Cohen99,LimMuniandy00}

MBM does not satisfy the self-similar property and its increments are no longer stationary as a result of the time-dependence of the H\"{o}lder exponent. 
However, one expects $B_{H(t)}$ to behave like FBM locally. 
If an additional condition is imposed on $H(t)$ such that 
$H(t) \in C^r\bigl(\mathbb{R},(0,1)\bigr), t \in \mathbb{R}$ for some positive 
$r$ with $r > \sup{H(t)}$, then it can be shown that $H(t_\circ)$ is almost certainly the H\"{o}lder exponent of the MBM at 
the point $t_\circ$; 
and the local Hausdorff and box dimensions of the graph of $B_{H(t)}$ at $t_\circ$ are almost certainly 
$2 - H(t_\circ)$. 
One can also characterize the above local fractal property by using the following notion. 
A process $Z(t)$ is said to satisfy the locally asymptotically self-similarity at a point $t_\circ$ if 
\begin{align}
  \lim_{\rho \to 0^{+}} \biggl[\frac{Z(t+\rho{u}) - Z(t_\circ)}{\rho^{H(t_\circ)}}\biggr]_{u \in \mathbb{R}} 
& = \bigl(B_{H(t_\circ)}(u)\bigr)_{u \in \mathbb{R}} ,
\label{eq:multifracBM_0010}
\end{align}
where the equality is up to a multipltcative deterministic function of time. 
It can be verified that $B_H(t)$ is locally asymptotically self-similar. 
Thus MBM at a time $t_\circ$ behaves locally like a FBM with H\"{o}lder exponent $H(t_\circ)$. 
Note that the time-dependent H\"{o}lder exponent
has no effect on the long range dependence of the process. 
Just like FBM, $B_{H(t)}$ is a long memory process.

Similar to the case of standard FBM, one can also extend FBM of RL type to its corresponding multifractional process. 
By replacing $H$ by $H(t)$ in (\ref{eq:fracBM_0110}),
one gets MBM of RL type with the following covariance:\cite{LimMuniandy00,Lim01}
\begin{align}
  \Bigl<X_{H(s)}X_{H(t)}\Bigr> & = \frac{{_2F_1\Bigl(1,\frac{1}{2} - H(t),H(s) + \frac{3}{2}, \frac{s}{t}\Bigr)}}
                                     { \bigl(2H(s)+1\bigr)
                                       \Gamma\Bigl(H(s)+\frac{1}{2}\Bigr)
                                       \Gamma\Bigl(H(t)+\frac{1}{2}\Bigr)
                                     }                                    
                                     s^{H(s)+\frac{1}{2}}t^{H(s)+\frac{1}{2}} .
\label{eq:multifracBM_0020}
\end{align}

The two type of MBM (standard and RL) have more properties in common as compared with the corresponding 
two types of FBM. 
They have non-stationary increments, and both are locally asymptotically self-similar with local fractal dimension at a point 
$t_\circ$ given by $2 - H(t_\circ)$, and they are both LRD.

MBM has been applied to model many phenomena which have variable irregurities or variable memory. 
For examples, it is used in modelling network traffic and signal processing,%
\cite{Krishna03,Sheng12ch6}
in geophysics for terrain modelling,%
\cite{Gaci12,Echelard10}
in financial time series for stochastic volatility modelling,%
\cite{Corlay14}
and in modelling anomalous diffusion with variable memory.%
\cite{LimMuniandy02,MarguezLago12}

Finally we remark that MBM can be further generalized. 
Various generalizations of to MBM have been proposed 
\cite{Ayache00,Vehel13}
to allow H\"{o}lder function to be very irregular, and enable the prescription of local intensity of jumps in space or time.


\section{Fractional and Multifractional Ornstein-Uhlenbeck Process }
\label{sec:mfracOU}
FBM and MBM are used to model long memory phenomena. For describing systems which are short-range dependent, Ornstein-Uhlenbeck process can be a suitable candidate. Recall that Ornstein-Uhlenbeck process is the solution ofthe ordinary Langevin equation 
\begin{align}
  D_t x(t) + \omega x(t) & = \eta(t) ,
\label{eq:mfracOU_0010}
\end{align}
where $\eta(t)$ is standard white noise
which can be regarded as the time derivative of Brownian motion in the sense of generalized function. 
Assuming $x(a) = 0$, the solution of (\ref{eq:mfracOU_0010}) is given by 
\begin{align}
  x(t) & = \int_a^t e^{\omega(t-u)} \eta(u)du .
\label{eq:mfracOU_0020}
\end{align}
There are several ways to fractionalize Ornstein-Uhlenbeck process. 
One way is to replace the white noise by a fractional Gaussian noise in 
(\ref{eq:mfracOU_0010})
or (\ref{eq:mfracOU_0020}),%
\cite{Cheridito03,LimMuniandy03}
or one can apply the Lamperti transformation to fractional Brownian motion.%
\cite{LimMuniandy03,Magdziarz08}

In this paper we shall consider a different type of FOU processes. 
FOU process of Weyl type and Riemann-Liouville type can be defined as\cite{LimEab06}
\begin{align}
  Y_\alpha^W(t) & = \frac{1}{\Gamma(\alpha)}\int_{-\infty}^t (t - u)^{\alpha-1} e^{\omega(t-u)} \eta(u)du , \label{eq:mfracOU_0030}\\
  Y_\alpha^{RL}(t) & = \frac{1}{\Gamma(\alpha)}\int_0^t (t - u)^{\alpha-1} e^{\omega(t-u)} \eta(u)du .
\label{eq:mfracOU_0040}
\end{align}
The condition $\alpha > 1/2$ is imposed to ensure finite variance for both the FOU processes. 
(\ref{eq:mfracOU_0030}) and (\ref{eq:mfracOU_0040}) can be regarded as the generalizations of (\ref{eq:mfracOU_0020}), 
with $a = -\infty$ and $a = 0$. 
These fractional processes are solutions to the following nonlinear fractional Langevin equation: 
\begin{align}
  \bigl({_aD_t + \omega}\bigr)^\alpha Y(t) & = \eta(t) .
\label{eq:mfracOU_0050}
\end{align}

The Weyl fractional Ornstein-Uhlenbeck process $Y_\alpha^W(t)$ is stationary 
centred Gaussian process with variance and covariance
\begin{subequations}
\label{eq:mfracOU_x010}
\begin{align}
  E\Bigl(\bigl[Y_\alpha^W(t)\bigr]^2\Bigr) & = \frac{\Gamma(2\alpha-1)(2\omega)^{1-2\alpha}}{\Gamma(\alpha)^2}, 
\label{eq:mfracOU_x010a}\\
  E\Bigl(Y_\alpha^W(t)Y_\alpha^W(s)\Bigr) & = \frac{1}{\sqrt{\pi}\Gamma(\alpha)}
                                           \left(\frac{|t-s|}{2\omega}\right)^{\alpha-1/2}
                                           K_{\alpha-1/2}\bigl(\omega|t - s|\bigr), &
                                           t & \neq s,
\label{eq:mfracOU_x010b}
\end{align}  
\end{subequations}
where $K_\nu(z)$ is the modified Bessel function of second kind.\cite{LimTeo09b}
On the other hand, the Riemann-Liouville fractional Ornstein-Uhlenbeck process 
$Y_\alpha^{RL}(t)$ is a non-stationary centred Gaussian process with variance and covariance 
\begin{subequations}
\label{eq:mfracOU_x020}
  \begin{align}
     E\Bigl(\bigl[Y_\alpha^{RL}(t)\bigr]^2\Bigr) & = \frac{(2\omega)^{1-2\alpha}\gamma(2\alpha-1,2\omega{t})}{\Gamma(\alpha)^2},  
\label{eq:mfracOU_x020a}\\
     E\Bigl(Y_\alpha^{RL}(t)Y_\alpha^{RL}(s)\Bigr) & = \frac{e^{-\omega(t+s)}s^\alpha t^{\alpha-1}}{\Gamma(\alpha+1)\Gamma(\alpha)}
                                                   \Phi_1\Bigl(1,1-\alpha,1+\alpha),\frac{s}{t},2\omega(s)\Bigr), 
                                               \ t  > s,                                                   
\label{eq:mfracOU_x020b}
  \end{align}
\end{subequations}
where $\gamma(a,x)$ is the incomplete Gamma function, and 
$\Phi_1(a,b,c,x,y)$ is the confluent hypergeometric function in two variables.
For discussion of properties and applications of the FOU process of Weyl and RL type, and their extension to FOU process with two indices 
can be found elsewhere.\cite{LimEab06,LimLiTeo08,LimTeo09}

Just like the case of MBM, one can extend the two types of FOU processes to their corresponding multifractional OU (MOU) 
processes by replacing $\alpha$ by $\alpha(t)$. 
The covariance of the MOU process of Weyl type for $s < t$ is given by
\begin{gather}
  E\Bigl(Y_{\alpha(t)}^W(t)Y_{\alpha(s)}^W(s)\Bigr)  = \frac{e^{-\omega(t+s)}}{\Gamma\bigl(\alpha(t)\bigr)\Gamma\bigl(\alpha(s)\bigr)}
                                                    \int_{-\infty}^s (t - u)^{\alpha(t)-1}(s - u)^{\alpha(s)-1} e^{2\omega{u}}du \nonumber \\
                                                 = \frac{e^{-\omega(t-s)}}{\Gamma\bigl(\alpha(t)\bigr)\Gamma\bigl(\alpha(s)\bigr)}
                                                    \int_0^\infty u^{\alpha(s)-1}(u+ t - s)^{\alpha(t)-1} e^{-2\omega{u}}du \nonumber \\
                                          \qquad       = \frac{e^{-\omega(t-s)}(t-s)^{\alpha(s)+\alpha(t)-1}}{\Gamma\bigl(\alpha(t)\bigr)}
                                                    \Psi\bigl(\alpha(s),\alpha(s) + \alpha(t), 2\omega(t-s)\bigr),
\label{eq:mfracOU_0060}
\end{gather}
where 
$\Psi(\alpha,y;z)$ is the confluent hypergeometric function. 
In contrast to the Weyl fractional Ornstein-Uhlenbeck process, the multifractional process is in general not stationary. 

For MOU process of RL type, its covariance for $s < t$ is
\begin{gather}
  E\Bigl(Y_{\alpha(t)}^{RL}(t)Y_{\alpha(s)}^{RL}(s)\Bigr)  = \frac{e^{-\omega(t+s)}}{\Gamma\bigl(\alpha(t)\bigr)\Gamma\bigl(\alpha(s)\bigr)}
                                                         \int_0^s (t - u)^{\alpha(t)-1}(s - u)^{\alpha(s)-1} e^{2\omega{u}}du \nonumber \\
                                                     = \frac{e^{-\omega(t+s)}s^{\alpha(s)}t^{\alpha(t)-1}}{\Gamma\bigl(\alpha(t)\bigr)\Gamma\bigl(\alpha(s)\bigr)}
                                                    \int_0^1 (1 - u)^{\alpha(s)-1}\Bigl(1 - \frac{s}{t}u\Bigr)^{\alpha(t)-1} e^{2\omega{us}}du \nonumber \\
                                           \mkern-27mu 
          = \frac{e^{-\omega(t+s)}s^{\alpha(s)}t^{\alpha(t)-1}}{\Gamma\bigl(\alpha(s)+1\bigr)\Gamma\bigl(\alpha(s)\bigr)}
                                                       \Phi_1\bigl(1,1-\alpha(t),1+\alpha(s),s/t,2\omega(s)\bigr).
\label{eq:mfracOU_0070}
\end{gather}

The local properties of these two types of MOU processes are similar to that of the corresponding MBM. 
With probability one, both the functions $Y_\alpha^W(t)$ and $Y_\alpha^{RL}(t)$  have 
H\"{o}lder exponent $\alpha(t_\circ) - 1/2$  at the point $t_\circ$; and the Hausdorff dimension of the two processes is $5/2 - \alpha(t_\circ)$.\cite{LimTeo07}
In addition, MOU processes of Weyl and RL-type are locally asymptotically self-similar, 
their tangent process at a point $t_\circ$ is the FBM indexed by parameter $\alpha(t_\circ) - 1/2$. 
In contrast to MBM, MOU processes are SRD, that is they are short memory processes.

The remark concerning the multifractality of the multifractional Brownian motion applies to the multifractional Ornstein-Uhlenbeck process. 
That is, the multifractional process is truly multifractal if the H\"{o}lder exponent is a random function, otherwise it is a multiscaling process. 
However, there are many phenomena that are multiscaling instead of multifractal.


\section{Fractional and Multifractional Riesz-Bessel Motion}
\label{sec:mfracRB}

Fractional Riesz-Bessel motion (FRBM) was first introduced by Anh et al. as fractional Riesz-Bessel random field.\cite{Anh99} 
In one dimension, it is a Gaussian process parametrized by two indices which characterize separately two distinct properties --- self-similarity and intermittency. 
The latter property corresponds to features such as sharp peaks or random bursts, and properties of processes 
that can be described by high skewed probability distributions with very slowly decaying tails. 
Thus FRBM has an advantage over FBM, which is unable to describe intermittency. 
In addition, for certain ranges of the two parameters, FRBM has a semimartigale representation.\cite{AnhLeonenkoMcvinish01}

FRBM is closely related to Riesz and Bessel potentials. 
In the one dimension case, FRBM can be regarded as the solution of the following fractional stochastic differential equation:
\begin{align}
  D_t^{\gamma/2}\bigl(D_t + \omega\bigr)^{\alpha/2} V_{\alpha,\gamma} & = \eta(t), &
                                                        \alpha & \geq 0, \     
                                                            0  \leq \gamma < 1 ,
\label{eq:mfracRB_0010}
\end{align}
where $D_t^{\gamma/2}$ is the Riesz derivative defined by
\begin{align}
  D_t^{\gamma/2}f(t) & = F^{-1}\Bigl(|k|^\gamma \hat{f}(k)\Bigr),
\label{eq:mfracRB_0020}
\end{align}
where $F$ denotes Fourier transform, $\hat{f} = F(f)$.
Formally, the solution of (\ref{eq:mfracRB_0010}) is given by
\begin{align}
  V_{\alpha,\gamma}(t) & = \frac{1}{2\pi} \int_{\mathbb{R}} \frac{e^{ikt}}{|k|^\gamma \bigl(\omega^2 + k^2\bigr)^{\alpha/2}} \eta(t)dt.
\label{eq:mfracRB_0030}
\end{align}
(\ref{eq:mfracRB_0030}) is to be regarded as a generalized random process.

Note that $V_{\alpha,\gamma}(t)$ can be defined as an ordinary stochastic process 
if $0 \leq \gamma < 1/2$, and $\alpha+\gamma > 1/2$. 
In the limit $\gamma = 0$, 
$V_{\alpha,\gamma}(t)$ becomes FOU process of Weyl type which is SRD.
On the other hand, if $\alpha = 0$,  (\ref{eq:mfracRB_0030})
becomes the generalized spectral density associated with FBM, a long memory process. 
In general, $V_{\alpha,\gamma}(t)$ is LRD when $\gamma \neq 0$.  
Thus, FRBM allows interpolation between long and short memory processes.

The spectral density of $V_{\alpha,\gamma}(t)$  is
\begin{align}
  S(k) & =  \frac{1}{(2\pi)|k|^{2\gamma} \bigl(\omega^2 + k^2\bigr)^{\alpha}} 
\label{eq:mfracRB_0040}
\end{align}
The covariance of FRBM can be calculated as the inverse Fourier transform of the spectral density (\ref{eq:mfracRB_0040})
\begin{align}
  C_{\alpha,\gamma}(x) & = \frac{\omega^{1-2\alpha-2\gamma}\Gamma\Bigl(\frac{1}{2} - \gamma\Bigr)\Gamma\Bigl(\alpha+\gamma-\frac{1}{2}\Bigr)}
                             {2\pi\Gamma(\alpha)}
                             {_1\mspace{-2mu}F_2}
                             \biggl(
                               \frac{1}{2}- \gamma; 
                               \frac{3}{2} -\alpha -\gamma, \frac{1}{2}; 
                               \Bigl[\frac{\omega|x|}{2}\Bigr]^2
                             \biggr)  \nonumber \\
                    & \quad + \frac{|x|^{2\alpha+2\gamma-1} \Gamma\Bigl(\frac{1}{2} - \alpha - \gamma\Bigr)}
                                 {2^{2\alpha+2\gamma}\sqrt{\pi}\Gamma(\alpha+\gamma)}
                                 {_1\mspace{-2mu}F_2}
                                 \biggl(
                                   \alpha; 
                                   \alpha+\gamma, \alpha + \gamma + \frac{1}{2};
                                   \Bigl[\frac{\omega|x|}{2}\Bigr]^2
                                 \biggr),
\label{eq:mfracRB_0050}
\end{align}
Note that when $\gamma=0$, (\ref{eq:mfracRB_0050}) becomes
\begin{align}
  C_{\alpha,0}(x) & = \frac{2^{1/2-\alpha}}{\sqrt{\pi}\Gamma(\alpha)}
                    \left(\frac{|x|}{\omega}\right)^{\alpha-1/2} K_{\alpha-1/2}\bigl(\omega|x|\bigr)
\label{eq:mfracRB_0060}
\end{align}
which is the covariance of the fractional Bessel process.\cite{LimTeo09b} 
When $\alpha=1$, (\ref{eq:mfracRB_0060}) becomes the two-point Schwinger function of the one-dimensional Euclidean scalar massive field.

Additional properties of FRBM are discussed elsewhere.%
\cite{Anh99,AnhLeonenkoMcvinish01,LimTeo08}
Generalization of FRBM to multifractional RBM (MRBM) can again be carried out by replacing
$\alpha$  and $\gamma$ by $\alpha(t)$ and $\gamma(t)$ respectively in (\ref{eq:mfracRB_0030}). 
The resulting MRBM $V_{\alpha(t),\gamma(t)}(t)$ is a Gaussian process which has many properties similar to MBM.
For examples, MRBM is locally asymptotically self-similar, its tangent process at a point $t_\circ$  is a standard FBM indexed by
$\alpha(t_\circ) + \gamma(t_\circ) - 1/2$.
Note that this is an example of the general result of Falconer\cite{Falconer03} that under certain conditions, the tangent process of 
a Gaussian process is FBM up to a multiplicative deterministic function of time.
Another local property is that the Hausdorff dimension at a point $t_\circ$ of the graph of FRBM is with probability one equals to
$5/2 - \alpha(t_\circ) - \gamma(t_\circ)$.

Finally, we consider the LRD and SRD properties of MRBM. 
In the general case where $\alpha(t)$ and $\gamma(t)$   are not constants, we can show the following:\cite{LimTeo08}
(a). If $\gamma(t)=0$ and there exists a constant $M$ so that
$n/2 < \alpha(t) \leq M$ then the MRBM of variable order $V_{\alpha(t),0}(t)$  is SRD.
(b). If there exist constants  
$L_1 \in (0,n/2)$ and $L_2 > n/2$,
$M_1 \in (L_1,n/2)$, $M_2 \geq L_2$
so that $L_1 \leq \gamma(t) \leq M_1$ and $L_2 \leq \alpha(t)+\gamma(t) \leq M_2$,
then MRBM $V_{\alpha(t),\gamma(t)}(t)$  is LRD.

FRBM and MRBM can be used to model systems that exhibit both long-range dependence and intermittency. For examples, in financial time series, air pollution, rainfall data, porosity in heterogenous aquifer, turbulence, etc.%
\cite{Anh99,AnhLeonenkoMcvinish01,Gao02,Anh98,Gao04}


\section{Generalized Cauchy Process}
\label{sec:genCauchy}

The stationary Gaussian process defined by the following generalized Cauchy (GC) covariance parametrized by two indices
\begin{align}
  C_{\alpha,\beta}(t) & = \Bigl<U_{\alpha,\beta}(s)U_{\alpha,\beta}(t+s)\Bigr>
                    = \bigl(1+ |t|^\alpha\bigr)^{-\beta}, &
                  & t \in \mathbb{R}, \ 0 < \alpha \leq 1, \ \beta > 0,
\label{eq:genCauchy_0010}
\end{align}
is known as generalized Cauchy process.%
\cite{LimLi06}
When $\alpha = 2$, $\beta = 1$ one gets the usual Cauchy process.
This process was first introduced by Gneiting and Schlather.\cite{GneitingSchlather04}
It has a nice and useful property which allows separate characterization of fractal dimension and LRD by two different parameters.

It is well-known that a stationary process cannot be self-similar.   
$U_{\alpha,\beta}(t)$ satisfies a weaker self-similar property known as local self-similarity.%
\cite{Kent97,Adler81}
A Gaussian stationary process is locally self-similar of order $\kappa$ if its covariance $C(t)$ satisfies for 
$t \to 0$,
\begin{align}
  C(t) & = 1 - \beta|t|^\kappa \Bigl[1+ O\bigl(|t|^\nu\bigr)\Bigr], &
       & \nu > 0.
\label{eq:genCauchy_0020}
\end{align}
A more intuitive alternative definition is the following. 
A Gaussian process $U(t)$ is said to be locally self-similar of index $\kappa$ if
\begin{align}
  U_{\alpha,\beta}(s) - U_{\alpha,\beta}(rt) & = r^\kappa \Bigl[U_{\alpha,\beta}(s) - U_{\alpha,\beta}(t)\Bigr], &
                                      & \text{as} \ |t - s| \to 0,
\label{eq:genCauchy_0030}
\end{align}
where the equality is in the sense of finite joint distributions. 
The above two definitions and also the locally asymptotically self-similarity defined by
(\ref{eq:multifracBM_0010}) 
are all equivalent.\cite{LimLi06}
It is straight forward to show that the tangent process at a point $t_\circ$ is FBM indexed by $\alpha$. 
In other words, GC process behaves locally like a FBM. 
The fractal dimension of the graph of a locally self-similar process of order $\alpha$ is
$5/2 - \alpha$.

GC process is LRD for $0 < \alpha\beta \leq 1$ and is SRD if $\alpha\beta > 1$. 
The large time lag behaviour of the covariance 
(\ref{eq:genCauchy_0010}) is given by the hyperbolically decaying covariance 
$C(t) \sim |t|^{-\alpha\beta}$, $t \to \infty$ which is characteristic of LRD. 
If the covariance is re-expressed as 
$\bigl(1 + |t|^\alpha\bigr)^{-\zeta/\alpha}$
then the parameters $\alpha$ and $\zeta$, respectively, provide separate characterization of fractal dimension and LRD. 

It is interesting to point out that the covariance of GC process has the same functional form as the characteristic function of generalized Linnik distribution
\cite{LimTeo10}
and spectral density of the generalized Whittle-Mat\'{e}rn process.\cite{LimTeo09b}
There are also laws in physics which have this same analytic form. 
One example is the Havriliak-Negami relaxation law in the non-Debye relaxation theory.%
\cite{LimLi06}
Thus, all of these quantities should have the same analytic and asymptotic properties, 
and results obtained in any one of them are of relevance to the other. 

Applications of GC process can be found in geostatistics, telecommunication
and climate modelling.%
\cite{LimLi06,LiLim08,Schlather10}
Extension to GC field\cite{LimTeo09c}
is particularly useful for geological modeling.
Generalization of GC process to multifractional GC process so far has not been carried out. 
However, it is expected such a generalization would be similar to MRBM indexed by two variable parameters.


\section{Concluding remarks}
\label{sec:concludex}

From the brief discussion given above, one notes 
that many of the fractional and multifractional Gaussian processes have similar local properties, 
in particular the local self-similarity (or having FBM as the tangent process at a point). 
The LRD (or SRD) character is carried over from the fractional process to the corresponding multifractional process.
Some related processes such as step FBM
\cite{Benassi00}
and mixed FBM
\cite{Cheridito01}
are not included. The step FBM can be regarded as a special case of MBM, with
$H(t)$ a piecewise linear function. Such a multiscale process can be used to model
anomalous transport phenomena such as single-file diffusion.\cite{LimTeo09d}
Mixed FBM is a linear combination of two or more independent FBM, and it can be used to model
retarding anomalous diffusion,\cite{EabLim12}
financial time series,\cite{Mishura08,Cheridito01} telecommunication,\cite{Filatova08}
etc. As far as applications of fractional and multifractional stochastic processes are concerned, it is possible to select from a variety of processes one that provides the best description of the system under study. 

Finally, we remark that path integral formulation of fractional stochastic processes has recently attracted considerable interest from physicists as well as mathematicians.\cite{Wio13,Kleinert12,JanakiramanSebastian12,EabLim14}
In view of the fact that several candidate theories of quantum gravity
\cite{Modesto09,Ambjorn10,Calcagni10,Calcagni12}
share the idea that spacetime is multifractal, 
one would expect path integral formulation of fractional and multifractional stochastic processes may
play an important role in physics, just like the case in Brownian motion.


\section*{Acknowledgments}
S.C. Lim would like to thank the organizers of this workshop Chris and Victoria for the financial support and their hospitality during his stay in Jagna.

\appendix

\bibliographystyle{ws-ijmpcs}
\bibliography{JagnaFBMa}

\begin{thebibliography}{10}

\bibitem{OldhamSpanier74}
K.~B. Oldham and J.~Spanier, {\em The fractional calculus: theory and
  application of differentiation and integration to arbitrary order} (London:
  Academic Press, 1974).

\bibitem{MillerRoss93}
K.~S. Miller and B.~Ross, {\em An introduction to the fractional calculus and
  fractional differential equations} (New York: Wiley, 1993).

\bibitem{Ortigueira11}
M.~D. Ortigueira, {\em Fractional Calculus for Scientists and Engineers}
  (Springer, New York, 2011).

\bibitem{Herrmann13}
R.~Herrmann, {\em Fractional Calculus: An Introduction for Physicists .}, 2nd
  edn. (World Scientific, 2013).

\bibitem{Nigmatullin92}
R.~Nigmatullin, {\em Theor. Math. Phys.} {\bf 90}, 242  (1992).

\bibitem{Rutman95}
R.~S. Rutman, {\em Theor. Math. Phys.} {\bf 105}, 1509  (1995).

\bibitem{Tatom95}
F.~B. Tatom, {\em Fractals} {\bf 3}, 217  (1995).

\bibitem{oshrefiTorbatiHammond98}
M.~Moshrefi-Torbati and J.~K. Hammond, {\em J. Franklin Inst. B} {\bf 335}, p.
  1077–1086  (1998).

\bibitem{Podlubny02}
I.~Podlubny, {\em J. Fract. Calc. Appl. Anal.} {\bf 5}, 357  (2002).

\bibitem{TenreiroMachado03}
J.~A.~T. Machado, {\em J. Fract. Calc. Appl. Anal.} {\bf 6}, 73  (2003).

\bibitem{Stanislavsky04}
A.~A. Stanislavsky, {\em Theor. Math. Phys.} {\bf 138}, 418  (2004).

\bibitem{Samko93}
S.~G. Samko, A.~A. Kilbas and O.~I. Marichev, {\em Fractional integrals and
  derivatives: theory and applications} (New York: Gordon \& Breach, 1993).

\bibitem{Podlubny99}
I.~Podlubny, {\em Fractional differential equations} (San Diego: Academic
  Press, 1999).

\bibitem{Kilbas06}
A.~A. Kilbas, H.~M. Srivastava and J.~J. Trujillo, {\em Theory and Applications
  of Fractional Differential Equations} (Elsevier Science \& Technology,
  Amsterdam, 2006).

\bibitem{Diethelm10}
K.~Diethelm, {\em The Analysis of Fractional Differential Equations} (Springer,
  New York, 2010).

\bibitem{MetzlerKlafter00}
R.~Metzler and J.~Klafter, {\em Physics Reports} {\bf 339}, 1   (2000).

\bibitem{Hilfer00}
R.~Hilfer (ed.), {\em Applications of Fractional Calculus in Physics}
  (Singapore: World Scientific, 2000).

\bibitem{West03}
B.~J. West, M.~Bologna and P.~Grigolini, {\em Physics of Fractal Operators}
  (New York: Springer, 2003).

\bibitem{Zaslavsky05}
G.~M. Zaslavsky, {\em Hamiltonian Chaos and Fractional Dynamics} (Oxford:
  Oxford University, 2005).

\bibitem{Klages08}
R.~Klages, G.~Radons and I.~M. Sokolov (eds.), {\em Anomalous Transport;
  Foundations and Applications.} (Wiley-VCH, New York,, 2008).

\bibitem{Mainardi10}
F.~Mainardi, {\em Fractional Calculus and Waves in Linear Viscoelasticity: An
  Introduction to Mathematical Models} (Imperial College Press, London, 2010).

\bibitem{Das11}
S.~Das, {\em Functional Fractional Calculus for System Identification and
  Controls} (Springer, New York, 2011).

\bibitem{KlafterLimMetzler11}
J.~Klafter, S.~C. Lim and R.~Metzler (eds.), {\em Fractional Dynamics: Recent
  Advances} (World Scientific, Singapore, 2011).

\bibitem{Tarasov11}
V.~E. Tarasov, {\em Fractional Dynamics: Applications of Fractional Calculus to
  Dynamics of Particles, Fields and Media} (Springer, New York, 2011).

\bibitem{Diethelm12}
K.~Diethelm, D.~Baleanu and E.~Scalas, {\em Fractional Calculus: Models and
  Numerical Methods} (World Scientific, Singapore, 2012).

\bibitem{Sheng11}
H.~Sheng, Y.~Q. Chen and T.~S. Qiu, {\em Fractional Processes and
  Fractional-Order Signal Processing: Techniques and Applications} (Springer,
  New York, 2011).

\bibitem{Meerschaert12}
M.~M. Meerschaert and A.~Sikorskii, {\em Stochastic Models for Fractional
  Calculus} (De Gruyter, Boston, 2012).

\bibitem{Uchaikin12}
V.~Uchaikin and R.~Sibatov, {\em Fractional Kinetics in Solids: Anomalous
  Charge Transport in Semiconductors, Dielectrics and Nanosystems} (World
  Scientific, Singapore, 2012).

\bibitem{Atanackovic14}
T.~M. Atanackovic, S.~Pilipovic, B.~Stankovic and D.~Zorica, {\em Fractional
  Calculus with Applications in Mechanics: Wave Propagation, Impact and
  Variational Principles} (Wiley-ISTE, New York, 2014).

\bibitem{Biagini08}
F.~Biagini, Y.~Hu, B.~{\O{}}ksendal and T.~Zhang, {\em Stochastic Calculus for
  Fractional {B}rownian Motion and Applications} (Springer, New York, 2008).

\bibitem{Mishura08}
Y.~Mishura, {\em Stochastic Calculus for Fractional {B}rownian Motion and
  Related Processes} (Springer, New York, 2008).

\bibitem{Nourdin12}
I.~Nourdin, {\em Selected Aspects of Fractional {B}rownian Motion} (Springer,
  New York, 2012).

\bibitem{SamorodnitskyTaqqu94}
G.~Samorodnitsky and M.~S. Taqqu, {\em Stable Non-Gaussian Random Processes}
  (New York: Chapman and Hall, 1994).

\bibitem{Marquardt06}
T.~Marquardt, {\em Bernoulli} {\bf 12}, 1099  (2006).

\bibitem{Cohen13}
S.~Cohen, A.~Kuznetsov, A.~E. Kyprianou and V.~Rivero, {\em L\'{e}vy Matters
  II: Recent Progress in Theory and Applications: Fractional L\'{e}vy Fields,
  and Scale Functions} (Springer, New York, 2013).

\bibitem{Cheridito03}
P.~Cheridito, H.~Kawaguchi and M.~Maejima, {\em Electron. J. Probab.} {\bf 8},
  1  (2003).

\bibitem{LimMuniandy03}
S.~C. Lim and S.~V. Muniandy, {\em J. Phys. A} {\bf 36}, 3961  (2003).

\bibitem{Magdziarz08}
M.~Magdziarz, {\em Physica A} {\bf 387}, 123  (2008).

\bibitem{Russo06}
F.~Russo and C.~A. Tudor, {\em Stochastic Process. Appl.} {\bf 116}, 830
  (2006).

\bibitem{Cheridito01}
P.~Cheridito, {\em Bernoulli} {\bf 7}, 913  (2001).

\bibitem{LimLiTeo08}
S.~C. Lim, M.~Li and L.~P. Teo, {\em Phys. Lett. A} {\bf 372}, 6309  (2008).

\bibitem{LimTeo09}
S.~C. Lim and L.~P. Teo, {\em J. Phys. A: Math. Theor.} {\bf 42}, p. 065208
  (34pp)  (2009).

\bibitem{Anh99}
V.~Anh, J.~Angulo and M.~Ruiz-Medina, {\em J. Statist. Plann. Inference} {\bf
  80}, 95  (1999).

\bibitem{AnhLeonenkoMcvinish01}
V.~V. Anh, N.~N. Leonenko and R.~Mcvinish, {\em Fractals} {\bf 9}, 329  (2001).

\bibitem{GneitingSchlather04}
T.~Gneiting and M.~Schlather, {\em SIAM Rev.} {\bf 46}, 269  (2004).

\bibitem{LimLi06}
S.~Lim and M.~Li, {\em J. Phys. A: Math. Gen.} {\bf 39}, 2935  (2006).

\bibitem{PeltierVehel95}
R.~Peltier and J.~L{\'e}vy~V{\'e}hel, {\em {Multifractional Brownian Motion :
  Definition and Preliminary Results}}, Rapport de recherche RR-2645, INRIA
  (1995), 

\bibitem{BenassiJaffardRoux97}
A.~Benassi, S.~Jaffard and D.~Roux, {\em Rev. Mat. Iber.} {\bf 13}, 19  (1997).

\bibitem{LimTeo07}
S.~C. Lim and L.~P. Teo, {\em J. Phys. A: Math. Gen.} {\bf 40}, 6035  (2007).

\bibitem{RuizMedina03}
M.~D. Ruiz-Medina, V.~V. Anh and J.~M. Angulo, {\em Stochastic Anal. Appl.}
  {\bf 22}, 775  (2004).

\bibitem{LimTeo08}
S.~C. Lim and L.~P. Teo, {\em J. Math. Phys.} {\bf 49}, p. 013509  (2008).

\bibitem{MandelbrotVanNess68}
B.~B. Mandelbrot and J.~W. van Ness, {\em SIAM Rev.} {\bf 10}, 422  (1968).

\bibitem{BarnesAllan66}
J.~A. Barnes and D.~W. Allan, {\em Proc. IEEE} {\bf 54}, 170  (1966).

\bibitem{Sheluhin07}
O.~Sheluhin, S.~Smolskiy and A.~Osin, {\em Self-Similar Processes in
  Telecommunications} (John Wiley \& Sons, Southern Gate, Chichester, 2007).

\bibitem{Rostek09}
S.~Rostek, {\em Option Pricing in Fractional {B}rownian Markets} (Springer, New
  York, 2009).

\bibitem{KKK03a}
Y.~Kobelev, L.~Kobelev and Y.~Klimontovich, {\em Phys. Dokl.} {\bf 48}, 264
  (2003).

\bibitem{KKK03b}
Y.~Kobelev, L.~Kobelev and Y.~Klimontovich, {\em Phys. Dokl.} {\bf 48}, 285
  (2003).

\bibitem{Sun09}
H.~G. Sun, W.~Chen and Y.~Q. Chen, {\em Physica A} {\bf 388}, 4586   (2009).

\bibitem{Cohen99}
S.~Cohen, From self-similarity to local self-similarity: the estimation
  problem, in {\em Fractals\/},  eds. M.~Dekking, J.~L\'{e}vy-V\'{e}hel,
  E.~Lutton and C.~Tricot (Springer London, 1999), pp. 3--16.

\bibitem{LimMuniandy00}
S.~C. Lim and S.~V. Muniandy, {\em Phys. Lett. A} {\bf 266}, 140  (2000).

\bibitem{Lim01}
S.~C. Lim, {\em J. Phys. A: Math. Gen.} {\bf 34}, 1301  (2001).

\bibitem{Krishna03}
P.~M. Krishna, V.~M. Gadre and U.~B. Desai, {\em Multifractal Based Network
  Traffic Modeling} (Springer New York, 2003).

\bibitem{Sheng12ch6}
H.~Sheng, Y.~Q. Chen and T.~S. Qiu, {\em Fractional Processes and
  Fractional-Order Signal Processing: Techniques and Applications} (Springer,
  New York, 2012), ch.~6, pp. 149--160.

\bibitem{Gaci12}
S.~Gaci, J.~L\'{e}vy-V\'{e}hel, C.~Keylock and J.~W. and. D.~Schertzer (eds.),
  {\em Nonlinear Processes in Geophysics on Multifractional {B}rownian motions
  in geosciences} 2012.

\bibitem{Echelard10}
A.~Echelard, O.~Barri\'{e}re and J.~L\'{e}vy-V\'{e}hel, Terrain modeling with
  multifractional brownian motion and self-regulating processes, in {\em
  Computer Vision and Graphics\/}, , Lecture Notes in Computer Science
  Vol.~6374 (Springer Berlin Heidelberg, 2010), pp. 342--351.

\bibitem{Corlay14}
S.~Corlay, J.~Lebovits and J.~L. V\'{e}hel, {\em Math. Finance} {\bf 24}, 364
  (2014).

\bibitem{LimMuniandy02}
S.~C. Lim and S.~V. Muniandy, {\em Phys. Rev. E} {\bf 66}, p. 021114  (2002).

\bibitem{MarguezLago12}
T.~Marguez-Lago, A.~Leier and K.~Burrage, {\em IET Syst Biol} {\bf 6}, 134
  (2012).

\bibitem{Ayache00}
A.~Ayache and J.~L. V\'{e}hel, {\em Stat. Inference Stoch. Process.} {\bf 3}, 7
   (2000).

\bibitem{Vehel13}
J.~L. V\'{e}hel, {\em Nonlinear Process. Geophys.} {\bf 20}, 643  (2013).

\bibitem{LimEab06}
S.~C. Lim and C.~H. Eab, {\em Phys. Lett. A} {\bf 335}, 87  (2006).

\bibitem{LimTeo09b}
S.~C. Lim and L.~P. Teo, {\em J. Phys. A: Math. Theor.} {\bf 42}, p. 105202
  (21pp)  (2009).

\bibitem{Falconer03}
K.~J. Falconer, {\em J. Lond. Math. Soc.} {\bf 67}, 657  (2003).

\bibitem{Gao02}
J.~Gao, V.~V. Anh and C.~Heyde, {\em Stochastic Process. Appl.} {\bf 99}, 295
  (2002).

\bibitem{Anh98}
V.~V. Anh, J.~M. Angulo, M.~D. Ruiz-Medina and Q.~Tieng, {\em Environ. Model.
  Software} {\bf 13}, 233   (1998).

\bibitem{Gao04}
J.~Gao, {\em J. Appl. Probab.} {\bf 41}, 467  (2004).

\bibitem{Kent97}
J.~T. Kent and A.~T.~A. Wood, {\em J. R. Stat. Soc. B} {\bf 59}, 679  (1997).

\bibitem{Adler81}
A.~J. Adler, {\em The Geometry of Random Fields} (New York: Wiley, 1981).

\bibitem{LimTeo10}
S.~C. Lim and L.~P. Teo, {\em J. Fourier Anal. Appl.} {\bf 16}, 715  (2010).

\bibitem{LiLim08}
M.~Li and S.~C. Lim, {\em Physica A} {\bf 387}, 387, 2584  (2008).

\bibitem{Schlather10}
M.~Schlather, {\em Bernoulli} {\bf 16}, 780  (2010).

\bibitem{LimTeo09c}
S.~Lim and L.~Teo, {\em Stochastic Process. Appl.} {\bf 119}, 1325   (2009).

\bibitem{Benassi00}
A.~Benassi, P.~Bertrand, S.~Cohen and J.~Istas, {\em Stat. Inference Stoch.
  Process.} {\bf 3}, 101  (2000).

\bibitem{LimTeo09d}
S.~C. Lim and L.~P. Teo, {\em J. Stat. Mech.} {\bf 2009}, p. P08015  (2009).

\bibitem{EabLim12}
C.~H. Eab and S.~L. Lim, {\em J. Phys. A: Math. Theor.} {\bf 45}, p. 145001
  (2012).

\bibitem{Filatova08}
D.~Filatova, {\em Mixed fractional Brownian motion: some related questions for
  computer network traffic modelling}, in {\em International Conference on
  Signal and Electronic System\/},  2008, pp. 393--396.

\bibitem{Wio13}
H.~S. Wio, {\em Path Integrals for Stochastic Processes: An Introduction}
  (Singapore: World Scientific, 2013).
\newblock see also additional references quoted in the book.

\bibitem{Kleinert12}
H.~Kleinert, {\em Europhys. Lett.} {\bf 100}, p. 10001  (2012).

\bibitem{JanakiramanSebastian12}
D.~Janakiraman and K.~L. Sebastian, {\em Phys. Rev. E} {\bf 86}, p. 061105
  (2012).

\bibitem{EabLim14}
C.~H. Eab and S.~C. Lim, {\em preprint arXiv:1405.0653}   (2014).

\bibitem{Modesto09}
L.~Modesto, {\em Class. Quant. Grav.} {\bf 26}, p. 242002  (2009).

\bibitem{Ambjorn10}
J.~Ambj\o{}rn, A.~G\"{o}rlich, J.~Jurkiewicz and R.~Loll, {\em Phys. Lett. B}
  {\bf 690}, 420  (2010).

\bibitem{Calcagni10}
G.~Calcagni, {\em Phys. Rev. Lett.} {\bf 104}, p. 251301  (2010).

\bibitem{Calcagni12}
G.~Calcagni, {\em Journal of High Energy Physics} {\bf 2012:65}  (2012).

\end{thebibliography}

\end{document}